\documentclass[longauth]{aa} % for the long lists of affiliations 
\usepackage{graphicx}
\usepackage{txfonts}
\usepackage{fixltx2e}
\usepackage{color}
\usepackage{upgreek}
\usepackage{hyperref}
\usepackage{breakurl}
\def\src{PKS\,1510--089}
\def\fermi{\textit{Fermi}-LAT}

\begin{document}

\title{Multiwavelength observations of a VHE gamma-ray flare from \src\ in 2015}

% authors 05.10.2016  Format AA
%
\author{
M.~L.~Ahnen\inst{1} \and
S.~Ansoldi\inst{2,}\inst{25} \and
L.~A.~Antonelli\inst{3} \and
C.~Arcaro\inst{4} \and
A.~Babi\'c\inst{5} \and
B.~Banerjee\inst{6} \and
P.~Bangale\inst{7} \and
U.~Barres de Almeida\inst{7,}\inst{26} \and
J.~A.~Barrio\inst{8} \and
%J.~Becerra Gonz\'alez\inst{9,}\inst{10,}\inst{27} \and % moved to Fermi part
W.~Bednarek\inst{11}
% ...
}
\institute { ETH Zurich, CH-8093 Zurich, Switzerland
\and Universit\`a di Udine, and INFN Trieste, I-33100 Udine, Italy
\and INAF National Institute for Astrophysics, I-00136 Rome, Italy
\and Universit\`a di Padova and INFN, I-35131 Padova, Italy
\and Croatian MAGIC Consortium, Rudjer Boskovic Institute, University of Rijeka, University of Split - FESB, University of Zagreb - FER, University of Osijek,Croatia
\and Saha Institute of Nuclear Physics, 1/AF Bidhannagar, Salt Lake, Sector-1, Kolkata 700064, India
\and Max-Planck-Institut f\"ur Physik, D-80805 M\"unchen, Germany
\and Universidad Complutense, E-28040 Madrid, Spain
\and Inst. de Astrof\'isica de Canarias, E-38200 La Laguna, Tenerife, Spain
\and Universidad de La Laguna, Dpto. Astrof\'isica, E-38206 La Laguna, Tenerife, Spain
\and University of \L\'od\'z, PL-90236 Lodz, Poland
\and Deutsches Elektronen-Synchrotron (DESY), D-15738 Zeuthen, Germany
\newpage % paper style
\and Institut de Fisica d'Altes Energies (IFAE), The Barcelona Institute of Science and Technology, Campus UAB, 08193 Bellaterra (Barcelona), Spain
\and Universit\`a  di Siena, and INFN Pisa, I-53100 Siena, Italy
\and Universit\"at W\"urzburg, D-97074 W\"urzburg, Germany
\and Institute for Space Sciences (CSIC/IEEC), E-08193 Barcelona, Spain
\and Technische Universit\"at Dortmund, D-44221 Dortmund, Germany
\and Finnish MAGIC Consortium, Tuorla Observatory, University of Turku and Astronomy Division, University of Oulu, Finland
\and Unitat de F\'isica de les Radiacions, Departament de F\'isica, and CERES-IEEC, Universitat Aut\`onoma de Barcelona, E-08193 Bellaterra, Spain
%\newpage % referee style
\and Universitat de Barcelona, ICC, IEEC-UB, E-08028 Barcelona, Spain
\and Japanese MAGIC Consortium, ICRR, The University of Tokyo, Department of Physics and Hakubi Center, Kyoto University, Tokai University, The University of Tokushima, Japan
\and Inst. for Nucl. Research and Nucl. Energy, BG-1784 Sofia, Bulgaria
\and Universit\`a di Pisa, and INFN Pisa, I-56126 Pisa, Italy
\and ICREA and Institute for Space Sciences (CSIC/IEEC), E-08193 Barcelona, Spain
\and also at the Department of Physics of Kyoto University, Japan
\and now at Centro Brasileiro de Pesquisas F\'isicas (CBPF/MCTI), R. Dr. Xavier Sigaud, 150 - Urca, Rio de Janeiro - RJ, 22290-180, Brazil
% ...
}

\titlerunning{MWL observations of a VHE $\gamma$-ray flare from \src\ in 2015}
\authorrunning{Ahnen et al.}
\date{Received ; accepted }

% \abstract{}{}{}{}{} 
% 5 {} token are mandatory
 
  \abstract
   {\src\ is one of only a few flat spectrum radio quasars detected in the VHE (very-high-energy, $>100$\,GeV) gamma-ray band. 
}
   {
We study the broadband spectral and temporal properties of the \src\ emission during a high gamma-ray state.
}
   {We performed VHE gamma-ray observations of \src\ with the MAGIC telescopes during a long high gamma-ray state in May 2015. 
In order to perform broadband modelling of the source, we have also gathered contemporaneous multiwavelength data in radio, IR, optical photometry and polarization, UV, X-ray and GeV gamma-ray ranges.
We construct a broadband spectral energy distribution (SED) in two periods, selected according to VHE gamma-ray state.
}
   {\src\ has been detected by MAGIC during a few day-long observations performed in the middle of a long, high optical and gamma-ray state, showing for the first time a significant VHE gamma-ray variability. 
Similarly to the optical and gamma-ray high state of the source detected in 2012, it was accompanied by a rotation of the optical polarization angle and the emission of a new jet component observed in radio.
However, due to large uncertainty on the knot separation time, the association with the VHE gamma-ray emission cannot be firmly established. 
The spectral shape in the VHE band during the flare is similar to the ones obtained during previous measurements of the source.
The observed flux variability sets for the first time constraints on the size of the region from which VHE gamma rays are emitted.
We model the broadband SED in the framework of the external Compton scenario and discuss the possible emission site in view of multiwavelength data and alternative emission models.
}
   {}

%separated by --
   \keywords{galaxies: active – galaxies: jets – gamma rays: galaxies – quasars: individual: PKS 1510-089}

   \maketitle
%
%________________________________________________________________
\section{Introduction}
\src\ is a bright flat spectrum radio quasar (FSRQ) located at the redshift of $z=0.36$ \citep{ta96}.
The source is one of only six objects firmly classified as a FSRQ from which gamma-ray emission has been detected in the very-high-energy (VHE, $>100$\,GeV) range \citep{ab13}. 
Moreover, one of the highest recorded apparent speed of superluminal motion, up to $\sim 46\,c$, has been seen in the ultrarelativistic jet of \src\ \citep{jo05}.
Like in many other FSRQs, the GeV gamma-ray emission of \src\ is strongly variable \citep{ab10, sa13, al14}.
The doubling-time scales of the \src\ flares observed in the GeV range go down to 1\,h \citep{sa13}.

Most of the FSRQs have been detected in the VHE gamma-ray range during (usually short) flares (see, e.g., \citealp{al08,al11, ah15}).
Since 2013 MAGIC performs regular monitoring of \src .
Interestingly, until 2015, no variability was seen in \src\ in VHE gamma rays; neither in H.E.S.S \citep{ab13} nor in MAGIC \citep{al14} observations. 
One should note, however, that both VHE gamma-ray detections happened during long periods of enhanced optical and GeV gamma-ray activity.
Hence no low-state VHE gamma-ray emission has been established so far from \src .

In May 2015, a strong flare of \src\ was observed in GeV gamma-rays by the Large Area Telescope (LAT) on board the \textit{Fermi} satellite.
The source showed also at this time high activity in optical \citep{atel7799,atel7542} and IR bands \citep{atel7495, atel7804}. 
The high state triggered further MAGIC observations, which led to the detection of an enhanced VHE gamma-ray activity from the source \citep{atel7542}. 
The VHE gamma-ray emission has been also observed by the H.E.S.S. telescope \citep{za16}. 
In May 2016 another flare happened \citep{atel9102,atel9105}, with an even stronger VHE gamma-ray flux than in May 2015. 
The May 2016 flare  will be discussed in a separate paper.

In this paper we report on the observations of \src\ during the May 2015 flare.
In Section~\ref{sec:data} we shortly introduce the instruments which provided multiwavelength data  and describe the data reduction procedures.
In Section~\ref{sec:results} we present the multiwavelength behaviour of the source.
Section~\ref{sec:model} is devoted to the interpretation of the data in the framework of an external Compton model.
The most important results are summarized in Section~\ref{sec:conc}.

%__________________________________________________________________
\section{Instruments, observations and data analysis}\label{sec:data}

During the May 2015 outburst \src\ was observed by various instruments in a broad range of frequencies (from radio up to VHE gamma rays). 
In this section we introduce the different instruments and data sets and explain the data analysis procedure. 

\subsection{MAGIC}
MAGIC is a system of two Imaging Atmospheric Cherenkov Telescopes with a mirror dish diameter of 17\,m each. 
They are located in Canary Island of La Palma ($28.7^\circ$\,N, $17.9^\circ$\,W), at the height of 2200 m a.s.l. \citep{al16a}.
As \src\ is a southern source, only observable at zenith angle $>38^\circ$, the corresponding trigger threshold is $\gtrsim 90$\,GeV \citep{al16b}, about 1.7 times larger than for the low zenith observations. 

The MAGIC telescopes observed \src\ for 5.4 hours between 18 and 24 of May, 2015 (MJD 57160--57166).
The data have been analyzed using MARS, the standard analysis package of MAGIC \citep{za13, al16b}.
As part of the data set was affected by Calima\footnote{Calima is a dust wind originating in Saharian Air Layer.} we have applied a correction for the atmosphere transmission based on LIDAR information \citep{fg15}.

\subsection{\fermi}
\fermi\ monitors the gamma-ray sky every 3~h in the energy range from $20\,\mathrm{MeV}$ to beyond $300\,\mathrm{GeV}$ \citep{Atwood09}. 
An analysis of the (publicly available) Pass 8 SOURCE class events was performed for a Region of Interest (ROI) of $10^\circ$ radius centered at the position of \src . 
In order to reduce contamination from the Earth Limb, a zenith angle cut of $<90^\circ$ was applied. 
The analysis was performed with the ScienceTools software package version v10r0p5 using the \texttt{P8R2\_SOURCE\_V6}\footnote{
\burl{http://fermi.gsfc.nasa.gov/ssc/data/analysis/documentation/Cicerone/Cicerone_LAT_IRFs/IRF_overview.html}
} instrument response function and the \texttt{gll\_iem\_v06} and \texttt{iso\_P8R2\_SOURCE\_V6\_v06} models\footnote{
\burl{http://fermi.gsfc.nasa.gov/ssc/data/access/lat/BackgroundModels.html}} for the Galactic and isotropic diffuse emission \citep{ac16}, respectively.

An unbinned likelihood analysis was applied using \texttt{gtlike}, including in the model all 3FGL sources \citep{Acero15} within $20^\circ$ from \src . 
The spectral indices and fluxes were left free for sources within $10^\circ$, while sources from $10^\circ$ to $20^\circ$ have their parameters fixed to their catalog value. 
A first unbinned likelihood fit was performed for the events collected within almost 4 months of data from March 22, 2015 to July 19, 2015 (MJD 57103--57223) in the energy range between 100 MeV and 800 GeV. % 57103.5--57223
The sources with test statistic (TS; \citealp{ma96}) below 5 were removed from the model. 
Next, the optimized output model was used to produce the light curves and spectra of \src\ in different time bins (from 1 day to 3 hours) and energy ranges (E$>$100\,MeV, E$>$1\,GeV). 
For the calculation of the light curves, all sources were fixed in the model except \src\ for which both the flux normalization and the spectral index were left free and modeled as a power-law.
For the calculation of the spectral points, in addition also the spectral index of \src\ was fixed to its best fit value during the considered time period for which the spectral points are estimated. 
The normalization of the Galactic and isotropic diffuse emission models was left to vary freely during the calculation of both the light curves and the spectra.

\subsection{\textit{Swift}-XRT and UVOT}
The multi epochs (16 individual pointings) event-list obtained by the X-ray Telescope (\textit{XRT}) \citep{2004SPIE.5165..201B} on-board of \textit{Swift} satellite during the period of May 11 to May 25, 2015 (MJD 57153--57167) with the total exposure time of 26.6\,ks were downloaded from the publicly available database table SWIFTXRLOG (\textit{Swift-XRT} Instrument Log). %$\sim7.38$ h
The individual exposures ranged from 0.6 to 4\,ks. 
They were processed using the HEASOFT package version 6.18. 
All the observations from this period had been performed in photon counting (PC) mode. 
The source region was defined as a circle of 20 pixel ($\sim$47") radius at the center of the source, while the background region was defined by an annulus centered at the source with inner and outer radii of 40 ($\sim$94") and 80 pixels ($\sim$188"), respectively. 
The source and background spectra were extracted using \verb|XSELECT| task (v2.4c). 
The source spectrum count rate does not exceed 0.5 counts/s in any of the observation epochs. 
Therefore, no pile-up correction was needed.
For the light curve analysis we have combined 3 pairs of epochs (MJD 57157, 57161, 57163) separated by $\sim$2\,hr.

The \verb|xrtexpomap| task (v0.2.7) was used to correct the flux loss due to the fact that some of the CCD pixels were not used during the data collection. 
The \verb|xrtmkarf| task (v0.6.3) took into account vignetting and bad pixels. 
The \verb|grppha| task was used to group source spectra in a way that each bin contains 20 counts. 
\verb|XSPEC| task (v12.9.0i) was used to calculate the flux and power law model spectral parameters using fixed equivalent Galactic hydrogen column density of $n_H = 6.89 \times 10^{20}$ [cm$^{-2}$] \citep{2005A&A...440..775K}. 

The second instrument on board the \textit{Swift} satellite, the Ultraviolet/Optical Telescope (UVOT, \citealp{po08}) was used to monitor the flux of the source in the 180--600\,nm wavelength range.
Following \cite{ra10} we used an iterative procedure for the data calibration, where the effective wavelength, counts-to-flux conversion factor and Galactic extinction for each filter were calculated by taking into account the filter effective area and source spectral shape.
Out of the 16 pointings in the investigated time period, 8 were taken with a full set of filters (v, b, u, w1, m2, w2).
For the SED modelling we used pointings on MJD 57160  (with u and w2 filters) contemporaneous with the MAGIC flaring state, and on MJD 57165 (all filters available) during the post-flare MAGIC observations. 

\subsection{Optical photometry and polarization}
\src\ is regularly monitored as part of the Tuorla blazar monitoring program\footnote{\url{http://users.utu.fi/kani/1m}} in R-band using a 35\,cm Celestron telescope attached to the KVA (Kunglinga Vetenskapsakademi) telescope located at La Palma. 
The data analysis was performed with the semi-automatic pipeline using the standard analysis procedures (Nilsson et al. in prep). 
The differential photometry was performed using the comparison star magnitudes from \citet{villata97}. 
The magnitudes were corrected for the Galactic extinction using values from \citet{schlafy11}.

The optical polarization observations were performed with a number of instruments: Nordic Optical Telescope (NOT), Steward Observatory, Perkins Telescopes, RINGO3, AZT-8 and LX-200.
The NOT polarimetric observations were done with ALFOSC in the R-band using the standard setup for linear polarization observations (lambda/2 retarder followed by a calcite). 
The data were analyzed using the standard procedures with semi-automatic software as in \citet{hovatta16}.
The Steward polarimetric observations were obtained as part of an ongoing monitoring program of gamma-ray-bright blazars in support of the {\it Fermi} mission\footnote{\url{http://james.as.arizona.edu/~psmith/Fermi}}. 
The observations were performed in the 5000-7000\,\AA\, band.
The data analysis pipeline is described in \citet{smith}.
Polarimetric R-band observations were also provided by the 1.8\,m Perkins telescope at Lowell Observatory equipped with PRISM (Perkins Reimaging System).  
The data analysis was done following the standard procedures as in \cite{chatterjee08}.
Polarization observations were also taken with the RINGO3 polarimeter \citep{ar12} on the fully robotic and autonomous Liverpool Telescope on La Palma, Canary Islands \citep{st04} as part of the Liverpool blazar monitoring campaign (see \citealp{je16}), in collaboration with the Monitoring AGN Polarimetry at the LIverpool Telescope (MAPLIT) program. 
Simultaneous observations in the 'blue', 350– 640 nm; 'green', 650–760 nm; and 'red', 770– 1000 nm passbands were taken using a rapidly rotating (once per 4 seconds) polaroid which modulates the incoming beam of light in 8 rotor positions. 
For this work we use only the 650–760 nm measurements, which are the closest to the R-band used in the mentioned above polarization instruments. 
The beam is simultaneously split by 2 dichroic mirrors into three electron multiplying CCD cameras (EMCCDs). 
The combination of the flux from the 8 rotor positions using equations from \cite{cn02} can be used to find the linear Stokes parameters, which were used to calculate the degree and angle of polarization. 
Finaly, additional optical polarimetric data are reported here from the 70cm AZT-8 telescope (Crimea) and the 40cm LX-200 telescope (St.Petersburg), both equipped with nearly identical imaging  photometers-polarimeters~\citep{la08}. 
Polarimetric observations were performed using two Savart plates rotated by $45^\circ$ relative to each another. 
By swapping the plates, the observer can obtain the relative Stokes $q$ and $u$ parameters from the two split images of each source in the field.
Instrumental polarization was found via stars located near the object under the assumption that their radiation is unpolarized.
The electric vector position angle (EVPA) was corrected for the $n\times 180^\circ$ ambiguity by minimizing the difference to the closest data point unless there is a gap of over 14 days. 

\subsection{Infrared}
\src\ is monitored by a number of IR instruments.
We have used the publicly available data in B, V, R, J and K bands from the Small and Moderate Aperture Research Telescope System (SMARTS) instrument located at Cerro Tololo Interamerican Observatory (CTIO) in Chile.
The data reduction and calibration is described in \cite{bo12}.
We converted the magnitudes into flux units using \cite{be98} and corrected for the interstellar dust absorption following \citet{schlafy11}.

The observations at Teide Observatory (Canary Islands) were obtained with the 1.52\,m Carlos Sanchez Telescope (TCS), using the near-infrared camera CAIN during the nights of MJD 57162--57174.
This camera is equipped with a 256 $\times$ 256 pixels NICMOS-3 detector providing a scale of 1''/pixel.
Data were acquired in the three filters J, H and Ks. 
Observations were performed using a 5-point dither pattern (repeated twice) in order to facilitate a proper sky background subtraction. 
At each point, the exposure time was about 1 min, split in individual exposures of 10 s in the J filter and 6 s in the H and Ks filters to avoid saturation by sky brightness.
Image reduction was performed with the {\it caindr} package under the IRAF environment\footnote{Image Reduction and Analysis Facility, \burl{http://iraf.noao.edu/}}. 
Data reduction includes flat-fielding, sky subtraction, and the shift and combination of all frames taken in the same dither cycle. 
Photometric calibration was made based on field stars from the 2MASS catalogue \citep{cu03}. 
The photometric zero point was determined for each frame by averaging the offset between the instrumental and the 2MASS magnitudes of the catalogue. 
Deviant stars were excluded and typical errors remained below 5\%.

We have also used IR photometry data obtained with a 1.2m telescope of Mt Abu InfraRed Observatory (MIRO), India, mounted with the Near Infrared Camera and Spectrograph (NICS) equipped with 1024x1024 HgCdTe Hawaii array detector. 
The field of view is 8'$\times$8' with a pixel scale of 0.5''/pixel. 
The observations on \src\ were performed with a 4-position dither with offsets of 30 arcsec, keeping the comparison stars\footnote{\burl{https://www.lsw.uni-heidelberg.de/projects/extragalactic/charts/1510-089.html}} 1, 2, 3, 4 and 6  in the field of the source. 
The sky and dark contributions were removed using these dithered images and aperture photometry was performed using standard procedures under IRAF (see \cite{ba12} for details of data reduction and analysis). 
The source magnitudes in J, H and Ks bands were calibrated using correction factors obtained using weighted average of the standard values of comparison stars mentioned above.

\subsection{Radio}
Radio monitoring observations were performed with Mets\"ahovi Radio Telescope operating at 37 GHz frequency. 
The instrument and data reduction procedures are described in \citet{te98} and \citet{al14}.

\begin{figure}
\centering
\includegraphics[width=0.49\textwidth]{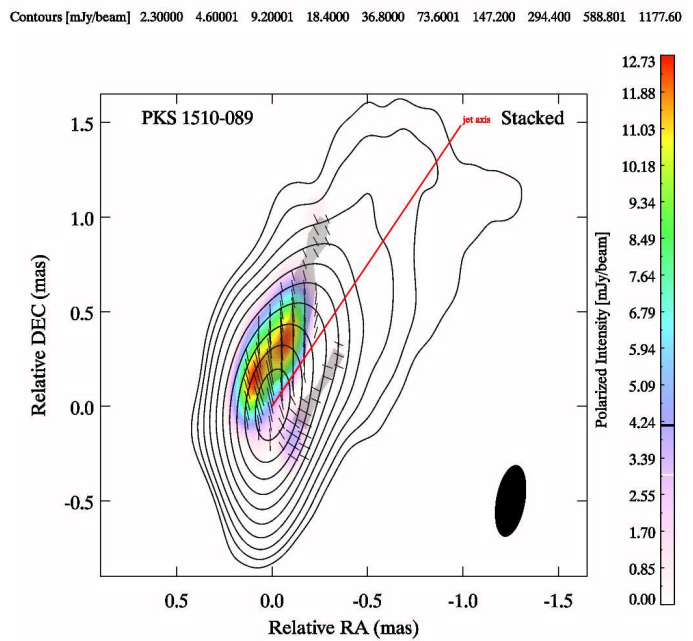}
\caption{
A stacked map of 43 GHz total (black contours) and polarized (color scale) intensity images of the inner, pc-scale jet of \src\ with the direction of electric field vector polarization denoted by black line segments (contour levels are indicated on the top, polarized  flux levels are shown by the  color bar to the right). All images have been convolved with the same Gaussian beam, shown in the lower right corner. 
}\label{fig:mapls}
\end{figure}

The quasar \src\ has been observed within a sample of gamma-ray blazars that the Boston University (BU) blazar group monitors with the Very Long Baseline Array (VLBA) approximately monthly at 43~GHz (the VLBA-BU-BLAZAR project). 
The observations of \src\ are usually performed during 9 short ($\sim 5$ min) scans within a span of 7-8 hours. 
The data were calibrated at the VLBA DiFX correlator and reduced using the Astronomical Image Process System (AIPS) and {\it Difmap} software packages, as described in \citet{jo05}. 
The calibrated data are available online (\url{www.bu.edu/blazars/VLBAproject.htm}). 
We analyzed the data obtained from  February 2015 to April 2016 (12 epochs). 
We modeled the total intensity images by components with circular gaussian brightness distributions. 
For each component we determined: flux density, distance and position angle (PA) with respect to the VLBI core\footnote{PA is measured starting from the positive direction of Declination axis (PA$=0^\circ$) increasing in the positive direction of Right Ascension axis (PA$=90^\circ$)}, and size. 
43 GHz core is expected to be located at the distance of $\sim$6.5\,pc from the central engine of \src\ (see \citealp{pu12, al14}).
A map of the parsec scale jet of the quasar formed from 20 stacked images over 6 years of VLBA observations at 43 GHz is plotted in Fig.~\ref{fig:mapls} (individual images can be found at the BU blazar group website \burl{http://www.bu.edu/blazars/VLBA_GLAST/1510.html})
The image shows the VLBI core, which is the brightest compact feature located at the southeast end of the jet.
The core is used as a reference point in the stacking procedure, since it is assumed to be stationary. 
The stacked image reveals the full opening angle of the jet, as well as the location of the jet axis. As can be inferred from Fig.~\ref{fig:mapls}, the jet axis is along PA$\sim -30\deg$, while the projected opening angle is $\sim 60\deg$. 
The core at 43 GHz is only partially optically thick, which is supported by synchronous optical and radio core polarization variability in a number of blazars \citep{da07,da09}. 
Fig.~\ref{fig:mapls} reveals a shift between the total and polarized intensity peaks and the complex structure of the polarized emission in the core. 
This favors the hypothesis that the core of \src\ is a recollimation shock, which has been inferred previously from the polarization structure in other blazars (e.g. \citealp{ca13}).

%__________________________________________________________________
\section{Results}\label{sec:results}

\begin{figure}
\centering
\includegraphics[width=0.49\textwidth]{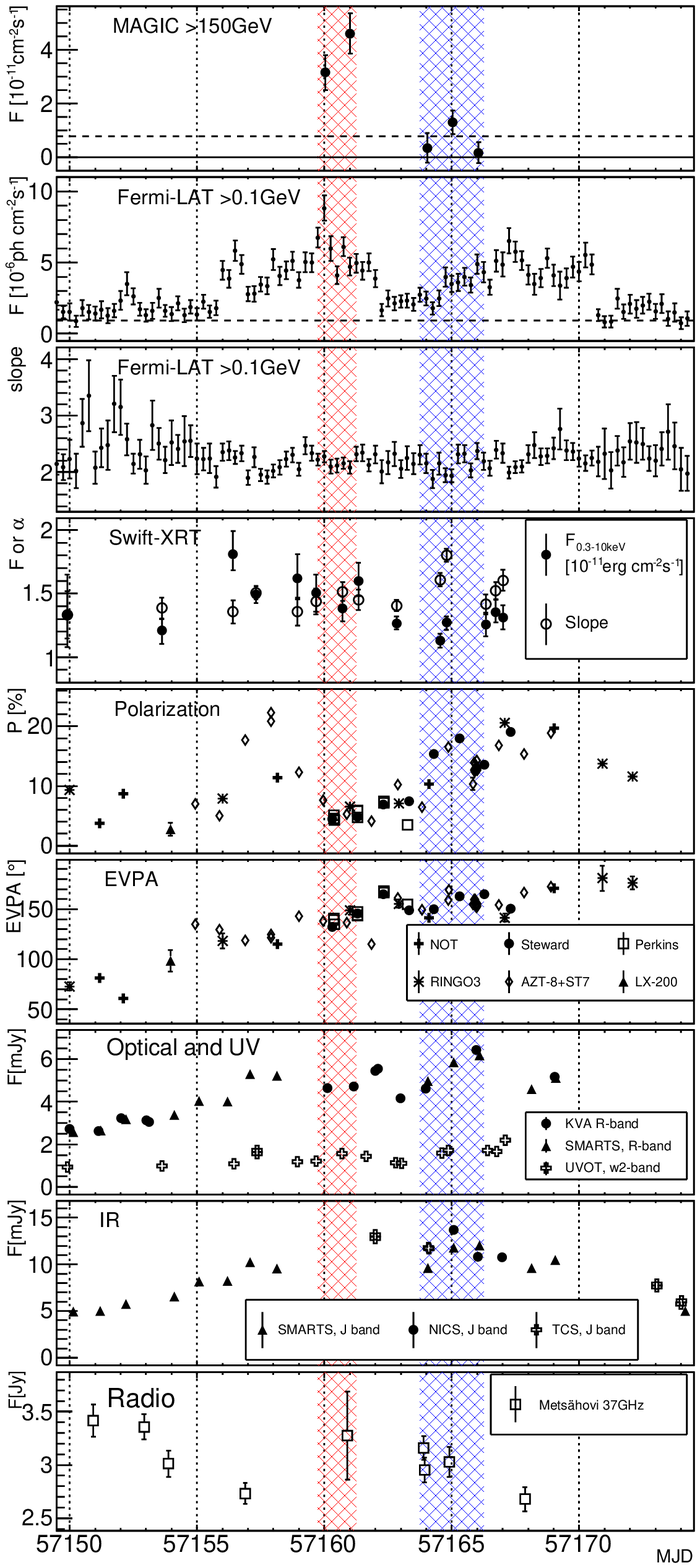}
\caption{
Multiwavelength light curve of \src\ during the May 2015 flare. 
From top to bottom: 
Nightly gamma-ray flux above 150\,GeV from MAGIC (the dashed line shows the average emission in Feb-Apr 2012, \citealp{al14});
\fermi\ flux above 0.1\,GeV in 6\,h binning, and the corresponding spectral index (the dashed line shows the average emission from the 3FGL catalog, \citealp{Acero15});
X-ray spectral flux (filled circles) and spectral index (empty circles) measured by \textit{Swift}-XRT;
polarization percentage and polarization angle measured by NOT, Steward, Perkins, RINGO3, AZT-8 and LX-200 (see legend);
optical emission in R band (KVA, SMARTS) and UV emission in w2-band (\textit{Swift}-UVOT);
IR emission in J band (SMARTS, MIRO-NICS, TCS);
radio observations by Mets\"ahovi at 37\,GHz. 
Data from IR up to UV are corrected for the Galactic absorption.
The red and blue shaded regions show the Period A and Period B, respectively, for which the spectral modelling is performed.
}\label{fig:mwllc}
\end{figure}

In Fig.~\ref{fig:mwllc} we show the multiwavelength light curve of \src\ during the May 2015 outburst in the investigated period of MJD 57151--57174.
Following the observations of MAGIC, we define two observation periods: Period A (MJD$\sim$ 57160-57161) and Period B (MJD$\sim$57164-57166). 
The multiwavelength SED of both periods is investigated. 

\subsection{MAGIC}
The MAGIC light curve (see top panel of Fig.~\ref{fig:mwllc}) shows clear variability, with the highest flux observed during the two nights of Period A. 
The hypothesis of constant flux during all 5 observation nights of MAGIC can be clearly rejected, with a chance probability of $7.7\times10^{-8}$. 
Even allowing for a 20\% variable systematic uncertainty on individual night fluxes (motivated by variable systematic uncertainty estimate given in \citealp{al16b}, rescaled to a softer source) we still obtain a small value of chance probability of $3.2\times10^{-4}$ for the flux to be constant. 
The flux during Period A is $\sim 5$ times larger than the one observed during the previous detection by MAGIC in 2012 \citep{al14}.
In the following observations during Period B, the VHE gamma-ray flux decreased to a level consistent with the detection in 2012. 

\begin{figure}
\centering
\includegraphics[width=0.49\textwidth]{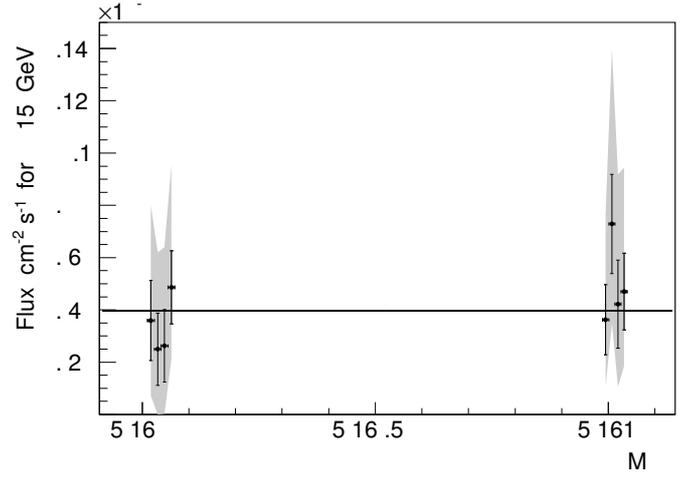}
\caption{
Light curve above 150\,GeV obtained with the MAGIC telescopes during the flare in Period A. 
The fluxes are computed in 20\,min bins. 
The black line shows the constant flux fit, $(4.0\pm0.5)\times 10^{-11} \mathrm{cm^{-2}s^{-1}}$.
The gray band shows the 95\% C.L. interval allowing for 20\% variable systematic uncertainty (see text for details).
}\label{fig:magiclc}
\end{figure}

In order to search for a possible short time variability we binned the light curve during the Period A into 20\,min bins (see Fig.~\ref{fig:magiclc}). 
No variability is detected at such a time scale.
Fitting the light curve with a constant flux hypothesis we obtain $\chi^2/N_{\rm dof} = 5.9/7$. 
We estimate the maximum variability which can be hidden by the uncertainties of the measurement by computing for each 20 min light curve bin a 95\% C.L. interval on the flux using \citet{ro05} prescription. 
We include a 20\% variable systematic uncertainty in those calculations.
By comparing the least constraining upper edge of the 95\% C.L. interval with the average flux from those two nights we obtained that the flux did not increase by more than a factor 3.5 on time scales of 20\,min. 

\begin{figure}
\centering
\includegraphics[width=0.49\textwidth]{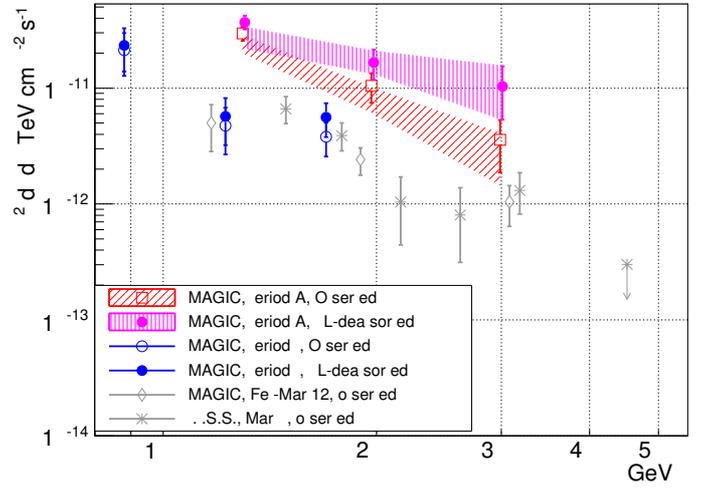}
\caption{
Spectral energy distribution of \src\ constructed from the MAGIC data gathered on MJD 57160 and 57161 (Period A). 
The observed spectrum is shown in red empty squares with inclined hashing, and the EBL-deabsorbed spectrum according to \citet{do11} model with magenta filled circles with vertical hashing. 
The hashing shows the uncertainty of the forward folding with a power law.
The SED constructed from the data taken on MJD between 57164 and 57166 (Period B) is shown as empty (observed) and filled (EBL-deabsorbed) blue circles. 
For comparison, MAGIC measurements performed in Feb-Mar 2012 and H.E.S.S. measurements from Mar 2009 \citep{ab13} are shown as gray diamonds and stars respectively.
}\label{fig:magicsed}
\end{figure}

For spectral analysis we have combined the two nights of Period A.
The obtained VHE gamma-ray spectrum of \src\ is shown in Fig.~\ref{fig:magicsed}.
It can be described by a power law,
$dN/dE=f \times (E/\mathrm{200\,GeV})^{-\alpha} \mathrm{[cm^{-2}\,s^{-1}\,TeV^{-1}]}$, with 
$f=(1.96\pm0.55_{\rm stat}\pm0.36_{\rm syst}) \times 10^{-10}$ and $\alpha=4.59\pm0.75_{\rm stat}$.
The spectral parameters are obtained  using a forward folding method \citep{al07}. 
The statistical uncertainty on the spectral index is much larger than the typical systematic uncertainty of $\pm 0.15$ of the observations performed with the MAGIC telescopes. 
The systematic uncertainty on the flux normalization does not include the uncertainty of the energy scale of MAGIC which for this data set we estimate as $\lesssim 19\%$, slightly larger than $\lesssim 15\%$ given in \cite{al16b}, due to the need of LIDAR correction of Calima-affected data. 
Correcting for the absorption of TeV gamma-rays due to the interaction with the extragalactic background light according to \citet{do11} model, an intrinsic spectrum with normalization of $f_{\rm deabs}=(4.2\pm1.0_{\rm stat}\pm0.76_{\rm syst}) \times 10^{-10}$ and index $\alpha_{\rm deabs}=3.17\pm0.80_{\rm stat}$ are obtained.
The spectral shape is marginally consistent (but with large uncertainties) with the previous measurements by H.E.S.S. (observed slope $5.4\pm0.7_{\rm stat}\pm0.3_{\rm syst}$, \citealp{ab13}) and  MAGIC (intrinsic slope $2.5\pm0.6_{\rm stat}$, \citealp{al14}).

For comparison, we have also reconstructed the average flux from the MAGIC measurements performed during Period B.
The combination of weaker emission and observations performed during higher atmospheric transmission results in the energy range of the reconstructed spectrum shifted to lower energies.
The observed flux in Period B is at a similar level as the one detected by MAGIC in 2012.
We obtained the observed and intrinsic VHE spectral slopes of \src\ during Period B of $4.75\pm0.62_{\rm stat}$ and $4.33\pm0.75_{\rm stat}$ respectively.

\subsection{\fermi}

The GeV gamma-ray flux of \src\ is highly variable in the investigated period. 
A few individual flares are visible, with time scales of a few days. 
Due to the short-term gamma-ray variability, in order to get a spectrum comparable to the MAGIC observations, for SED analysis we selected the events observed by the LAT within 6\,h centered in each of the MAGIC observations. 
We have calculated two GeV spectra, which correspond to the two different states of the source contemporaneously to the MAGIC observations in Period A and Period B, respectively. 
The best description of the GeV spectrum measured by the LAT in Period A is a power-law with spectral index 2.20$\pm$0.07 and a flux above 100 MeV of (6.8$\pm$0.5)$\times 10^{-6}\,\mathrm{ph}\,\mathrm{cm}^{-2} \mathrm{s}^{-1}$ ($\mathrm{TS}=842$). %842.1
During Period B, the \fermi\ measured spectrum is best described by a power-law with a similar spectral index of 2.17$\pm$0.08, but significantly lower flux, (3.7$\pm$0.3)$\times 10^{-6}\,\mathrm{ph}\,\mathrm{cm}^{-2} \mathrm{s}^{-1}$ ($\mathrm{TS}=564$). %563.9
For comparison, the 3FGL flux above 100 MeV is (0.94$\pm$0.01)$\times 10^{-6}\,\mathrm{ph}\,\mathrm{cm}^{-2} \mathrm{s}^{-1}$ \citep{Acero15}. 
Therefore \src\ has reached a factor of $\sim$ 7 to 4 times its average flux over the first 4 years of \fermi\ observations during the two epochs for which the spectral analyses has been performed. 
The GeV flux was, however, still a factor of about 2 smaller than the daily peak flux observed in 2011 \citep{sa15}. 
The \fermi\ spectrum during both Period A and B was slightly harder than in the neighbouring days (see also Section~\ref{sec:radio}). 
In the case of Period B there is a weak hint of a spectral curvature.
The likelihood ratio test gives $2.8\sigma$ preference ($\mathrm{\Delta TS=8.0}$ for one additional degree of freedom) of the log-parabola shape over the power-law spectral model. 
The corresponding photon index can be described as $(2.05\pm0.10)+(0.21\pm0.09)\times \ln(E/0.45\,\mathrm{GeV})$. %445 MeV

To characterize the variability in the \fermi\ light curve we calculated the Power Density Spectrum (PDS) both for the 2015 flare epoch (MJD 57143.9375--57182.9375) with a 3 h binning as well as for a mission lifetime at the time of the analysis (MJD 54682.655--57484.655) with 1 day binning. 
In both cases the flux calculation was performed in the 0.1-800 GeV energy range.

The estimated white noise level, based on the data error values was subtracted from each PDS. 
The PDS is rebinned in logarithmic frequency intervals, and normalized to variance per frequency unit divided by the square of the mean flux. 
\begin{figure}
\centering
\includegraphics[width=0.49\textwidth]{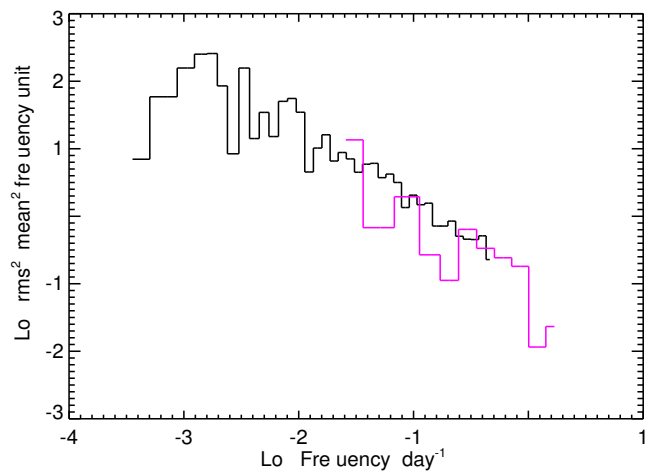}
\caption{Power Density Spectrum normalized to variance per frequency unit for the mission long \fermi\ light curve (black) and for the 2015 flare epoch (magenta).
}\label{fig:fermipds}
\end{figure}
The PDS level, computed with such normalization, is similar for the 2015 flare and for the mission lifetime light curve (see Fig.~\ref{fig:fermipds}). 
This suggests that the fractional variability is the same and presumably driven by the same variability process. 
The shape of the overall PDS can be described by a power law for frequencies above 0.01~day$^{-1}$. 
A power-law fitted to the 0.007-0.5~day$^{-1}$ frequency range PDS gives an index of 1.14\,$\pm$\,0.07 for full data set and 0.97\,$\pm$\,0.30 for the 2015 flare. 
The uncertainty value of the fit is based on the scatter of the measurements with respect to the fitted line only and therefore may be underestimated.
The power law index is similar to that of other FSRQs \citep{ac11}.
The stochastic nature of the variability together with the limited observation length lead to a large uncertainty in the PDS at the lowest frequencies.

\subsection{\textit{Swift}-XRT}

The X-ray flux, measured by \textit{Swift}-XRT, shows a gradual decrease of the observed flux during the studied period. 
Except for the first point at MJD 57153.6, which happened before the two large \fermi\ flares the X-ray flux can be much better described by an exponential decline ($\chi^2/N_\mathrm{dof}=22.6/10$) than a constant value ($\chi^2/N_\mathrm{dof}=47.3/11$).
Similarly, the corresponding X-ray spectral indicies are better described by a linear softening of the spectrum ($\chi^2/N_\mathrm{dof}=41.0/10$) than a constant ($\chi^2/N_\mathrm{dof}=49.9/11$).

The flux is marginally higher during  MJD 57156-57162, when a broad \fermi\ flare happened.
XRT observations on MJD 57166-57168, during the next broad \textit{Fermi} flare did not show a clear increase of the X-ray flux. 
For spectral analysis we combined the 4 pointings taken during Period A, each of them being 6-8\,hr distant from the MAGIC observations. 
The X-ray spectrum in Period A can be marginally well described ($\chi^2/N_{\rm dof}=71/53$) with a power law with an index of  $1.48 \pm 0.05$. %$1.478 \pm 0.047$. 
\textit{Swift}-XRT observations performed during Period B resulted in a spectrum which can be well fitted with a power law with a significantly softer index of $1.70\pm0.04$ with $\chi^2/N_{\rm dof}=110/89$, %1.703\pm0.036
or by a curved spectrum with an index of $2.23\pm0.14-(0.366\pm0.093)\times \log (E/0.3\mathrm{keV})$ with $\chi^2/N_{\rm dof}=96/88$.

\subsection{IR, Optical and UVOT}

The optical-UV SED of \src\ consists of an unpolarized, quasi-stable accretion disk component and non-thermal jet emission.
The variability and polarization of the jet component might be diluted by the accretion disk component. 
The effect is strongest in the UV bands and weak in the R-band, where the polarization is measured.   
It does not affect the timing of the observed polarization variability. 

The optical emission of \src\ during the investigated period is clearly variable (with a factor 2 difference between the lowest and highest flux).
The optical variability does not strictly follow the gamma-ray one. 
In the optical R-band, and to a lesser extent also in UV w2-band, the flux was slowly raising throughout May 2015. 
Similarly as in the optical range, the IR flux doubled in a $\sim8$ days time before Period A. 
At the end of the observation period it returned to the pre-flare state.

Throughout the investigated period, a smooth rotation of optical EVPA by $\sim 100^\circ$ is happening. 
The rotations of optical polarization angle accompanied also the 2009 and 2012 gamma-ray flaring states \citep{ma10,al14}.
However, the rotations of the EVPA seem very common in \src , e.g., recent work by \citet{je16} identified a rotation also in the 2011 data, therefore further data are needed to firmly associate them with the emission of VHE gamma rays.
The low percentage of polarization seems to be typical for this source \citep{je16}.
Indeed, during Period A the polarization percentage is low ($\sim5\%$).
It triples between the period A and B. 
Also in the few days before the Period A (i.e. during the raising part of the \fermi\ flare that culminated with detection of VHE gamma-ray emission by MAGIC) a higher polarization was observed. 
The polarization behaviour during the 2015 flaring period agrees with what one would expect from a knot following a spiral path through a mainly toroidal magnetic field \citep{ma10}. 
An alternative explanation might be the light travel time effects within an axisymmetric emission region pervaded by predominately helical magnetic field \citep{zh15}.
The evolution of the polarization is probably further complicated by superposition of individual flares seen in \fermi . 

\subsection{Radio}\label{sec:radio}

\src\ shows moderate variability in Mets\"ahovi observations performed at 37\,GHz in May 2015.
The sampling is, however, rather sparse, and especially the local peaks of the GeV flux are mostly not covered by the observations. 
The observations on MJD 57161, during Period A are burdened with a larger uncertainty due to adverse atmospheric conditions.

\begin{figure}[t]
\includegraphics[width=0.49\textwidth]{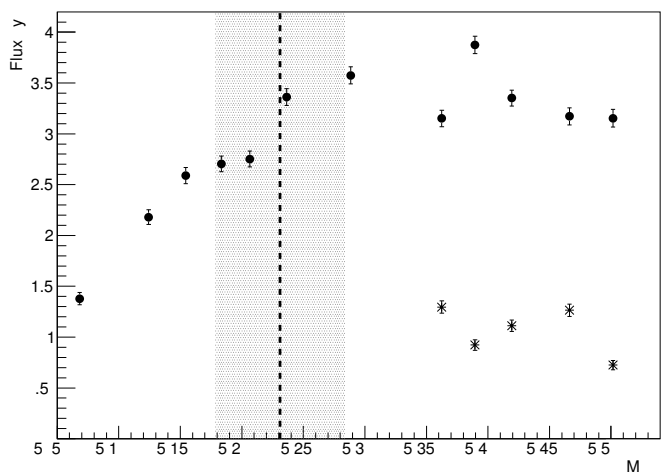}
\caption{
VLBA light curves of the core (circles) and knot $K15$ (stars) at 43 GHz.
Vertical dashed line and gray shaded region show the zero separation epoch of $K15$ and its uncertainty. 
}\label{fig:LCvlba}
\end{figure}

\begin{figure}[t!]
\centering
\includegraphics[width=0.4\textwidth]{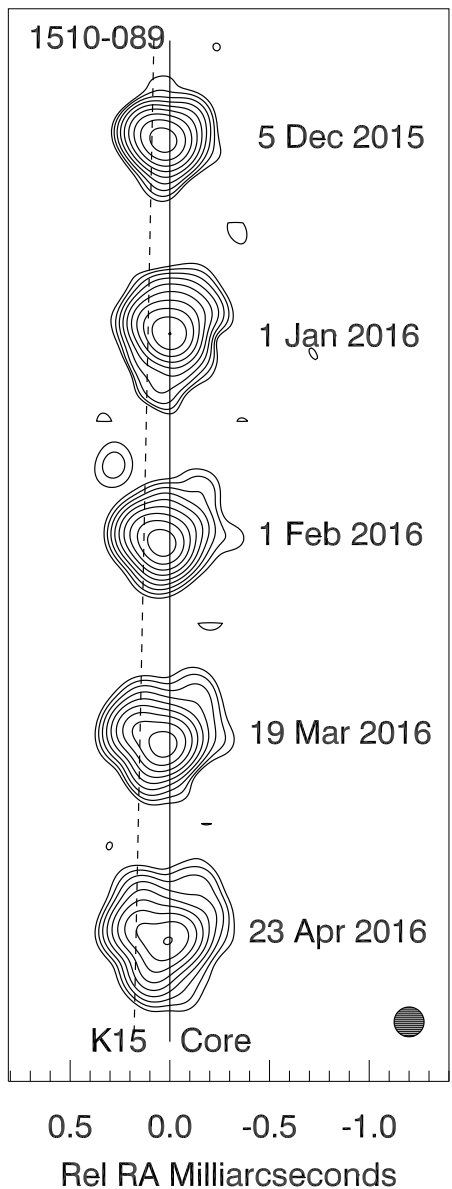}
\caption{
Total intensity images of PKS1510-089 core region at 43 GHz, with a global peak intensity of $I_{peak}=3.566$\,Jy/beam and 0.15~mas FWHM circular Gaussian restoring beam (bottom right circle). 
The solid and dashed lines designate positions of the VLBI core and $K15$ respectively across the epochs.
}\label{fig:1510t_vlba}
\end{figure}

\begin{figure}[t]
\includegraphics[width=0.49\textwidth]{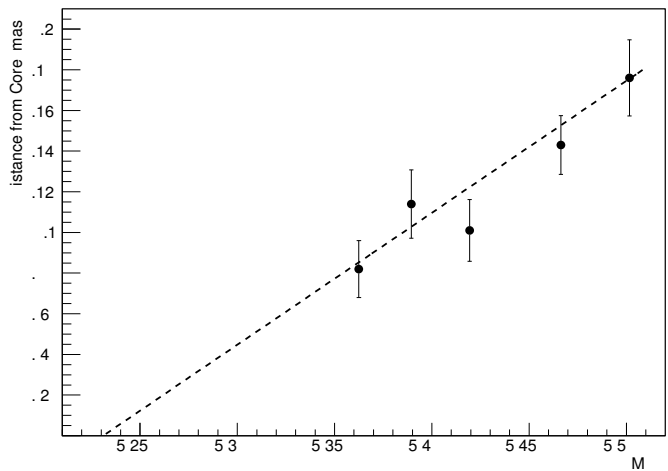}
\caption{
Separation of knot $K15$ from the core. 
}\label{fig:1510_move}
\end{figure}

\begin{figure}[t]
\includegraphics[width=0.49\textwidth]{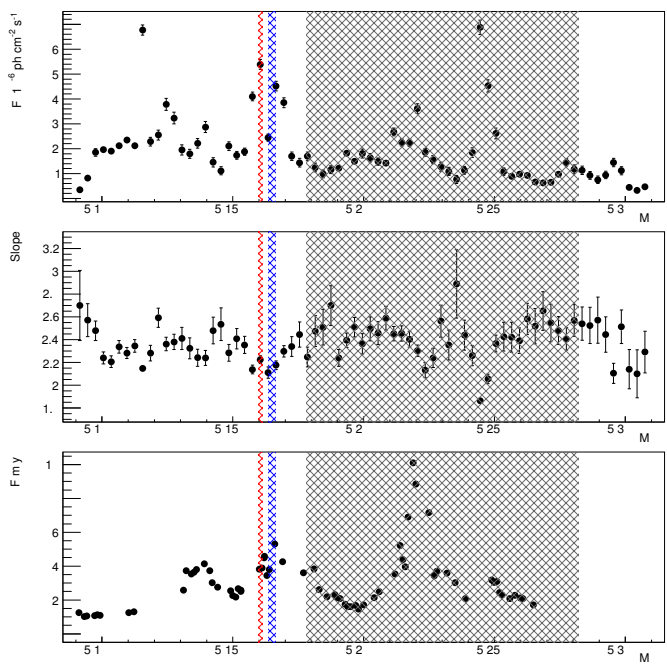}
\caption{
\fermi\ light curve above 0.1 GeV (top panel) binned in 3 day intervals and the corresponding evolution of spectral index (middle panel) and the optical  R-band flux observed by KVA (bottom panel).
Red and blue hashed bands show the Period A and B respectively, while the gray band shows the $1\sigma$ uncertainty on the $K15$ knot separation time. 
}\label{fig:fermilongterm}
\end{figure}

Fig.~\ref{fig:LCvlba} shows the light curve of the core from February 2015 to April 2016.
The light curve reveals a significant gradual increase of flux in the second half of 2015 that is most likely connected with a disturbance (knot $K15$) detected in the VLBA images starting in  December 2015 (Fig.~\ref{fig:1510t_vlba}). 
Knot $K15$ is bright and relatively slow, with an apparent speed $\beta_{app}$=5.3$\pm$1.4~c. %$\beta_{app}$=5.33$\pm$1.44~c. 
According to currently available data, the knot was ejected on MJD 57230$\pm$52 (see Fig.~\ref{fig:1510_move}). 
A similar behavior has also been observed during a high gamma-ray state in Feb-Apr 2012, when the emerging of a new radio knot ($K12$) from the core was associated with a VHE outburst \citep{al14}.
The large uncertainty in the ejection time of $K15$ does not allow us to associate it firmly with a particular GeV flare, as the source showed activity in \fermi\ close to the time of $K15$ separation (note e.g. the hard flare on MJD 57245 in Fig.~\ref{fig:fermilongterm}). 
Interestingly, the position angle of $K15$ is $\sim$+50$^\circ$, while the typical projected direction of the pc-scale jet is $\sim-$30$^\circ$ (see Fig.~\ref{fig:mapls}), along which so far 7 knots have been observed (see e.g. \citealp{jo05,al14}). 
On the other hand, knot $K11$, ejected in  October 2011 (before the 2012 VHE outburst), had a similar PA to the one observed now for $K15$ close to the core, but turned toward the usual pc-scale jet direction a few months later. 
We note that $K15$ is a factor of a few times slower than $K11$, and so may eventually follow a similar trajectory. 
The slower apparent speed and brighter flux of $K15$ suggest that the velocity vector of this disturbance is closer to the line of sight than $K11$, causing it to have a higher Doppler factor.

%__________________________________________________________________
\section{Modelling of the spectral energy distribution}\label{sec:model}

The gamma-ray emission of FSRQs is usually explained as the effect of the inverse Compton scattering of electrons on a radiation field external to the jet (see, e.g. \citealp{sbr94}; \citealp{gh10}), the so-called external Compton (EC) scenario.
The radiation field can originate from the accretion disk, broad line region (BLR) or the dust torus (DT). 
This scenario has been applied to explain the emission of \src\ in its previous flaring episodes \citep{ab10}.
The origin of the radiation field is closely connected with the location of the emission region.
Moreover, the observed VHE gamma-rays escaping from the emission region suggest that the emission region is located outside the BLR in order to escape the absorption by $e^+e^-$ pair production process on BLR photons \citep{ab13,al14}. 
\citet{do15} investigated the 2009 GeV flares of \src\ using the energy dependence of the flare decay time as the diagnostic for the emission zone location. 
They claimed that two of the flares happened around the distance of the DT, while the other two came from the vicinity of VLBI core. 
On the other hand, the modelling of GeV flares seen from \src\ in 2011 placed the emission region at the distance of $0.3-3$\,pc from the black hole, with the EC process happening on a mixture of BLR and DT radiation fields \citep{sa15}. 
A similar location ($\sim1$\,pc distance from the black hole) with EC mainly on DT was invoked to explain the high optical and gamma-ray state observed in the beginning of 2012 \citep{al14}. 
The broadband emission could be also explained by a much more distant ($\sim20$\,pc) region for a 2-zone model in which the jet consists of the inner spine and outer sheath layer \citep{al14,md15}. 

\begin{figure}[t]
\begin{minipage}{0.49\textwidth}
\includegraphics[width=\textwidth]{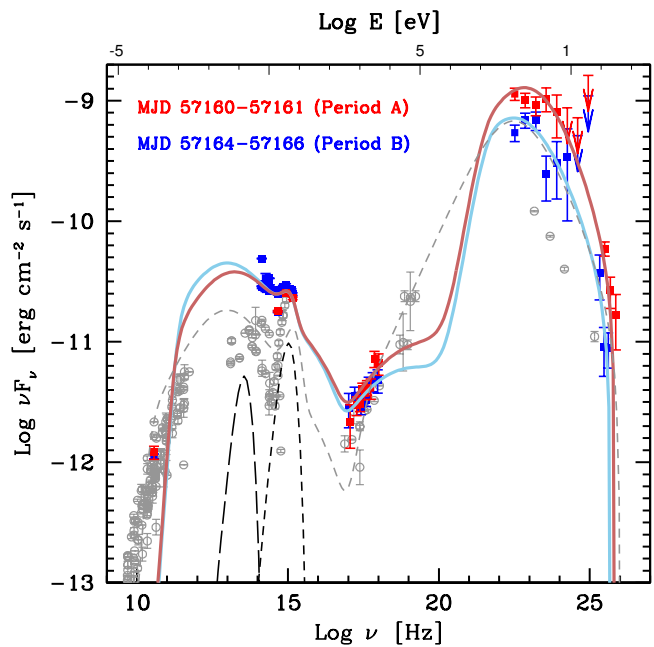}
\caption{Multiwavelength spectral energy distribution of \src\ in Period A (red symbols) and B (blue symbols). The red and the blue curves report the result of the emission model for the two periods. The black dashed and long-dashed lines show the adopted emission for the accretion disk and the dusty torus, respectively. For comparison, the dashed gray line shows the model derived for the SED in 2012 (from \citealt{al14}).
Historical measurements (ASDC, see \burl{http://www.asdc.asi.it/}) are shown as gray points.
}\label{fig:sed}
\end{minipage}
\end{figure}

In Fig.~\ref{fig:sed} we present the two spectral energy distributions of \src\ constructed from the data taken in Period A and Period B, corresponding to high and low gamma-ray flux, respectively. 
As can be seen, most of the flux variation ($\approx$ a factor of 2--3) occurs in the {\it Fermi}-LAT and MAGIC bands. 
The low-energy flux (optical, X-rays) is almost constant between the two periods. 
Remarkably, the high energy peak during the period B has a very similar level to the 2012 high state, in spite of the IR--UV emission being a factor of $\sim3$ higher.

We model these SEDs of \src\ in the framework of the same one-zone model as discussed in \cite{al14}.
To explain the sub-TeV emission observed by MAGIC we assume that the emission region is located beyond the BLR radius, where the external photon field is dominated by the thermal IR radiation of the DT. 

For the setup of the model we assume the scaling laws and the prescriptions given in \cite{gt09}. Specifically, the radius of the BLR is given by $R_{\rm BLR}=10^{17} L^{1/2}_{\rm d,45}$ cm and that of the torus by $R_{\rm IR}=2.5\times 10^{18} L^{1/2}_{\rm d,45}$ cm, where $L^{1/2}_{\rm d,45}$ is the accretion disk luminosity in units of $10^{45}$ erg s$^{-1}$. We calculate the IR radiation energy density assuming that a fraction $f_{\rm IR}=0.6$ of the disk radiation is intercepted and reprocessed by the torus heated to 1000\,K. 
Similarly, the BLR intercepts $f_{\rm BLR}=0.1$ of the disk radiation. 
Note that, with these prescriptions, the energy densities of the BLR and torus radiation fields do not depend on the disk luminosity, since they depends on the constant ratio $L_{\rm disk}/R_{\rm BLR\,or\,IR}^2$.
Assuming the same disk luminosity $L_{\rm disk}=6.7\times 10^{45}$ erg s$^{-1}$ as in \cite{al14} the scaling law of \cite{gt09} (based on reverberation mapping measurements of BLR size, see e.g. \citealp{be09}) allows us to infer a BLR radius of $R_{\rm BLR}=2.6\times 10^{17}$ cm. % (0.086 pc)

We fix the distance of the emission region from the base of the jet to $r=6\times 10^{17}$ cm.
The emission region is filling the whole cross section of the jet, which for an assumed jet semi-aperture angle $\theta_{\rm j}=0.047$\,rad, results in the emission region radius $R=2.8\times 10^{16}$ cm.
Such size of the emission region is in line, even for moderate values of the Doppler factor, with the constraints set by the few day time scale variability observed by MAGIC.

The apparent superluminal motion of radio component K15 puts limits on $\Gamma\gtrsim 5$. 
The large uncertainty in the separation time from the radio core does not allow us to firmly associate such limits with the speed of the emission region responsible for the emission in investigated periods A and B. 
Moreover, the apparent speed is measured over a much longer time scale, during which the emission region might have decelerated.
On the other hand the beaming of the emission is constrained by the observed luminosity of the dominating high energy peak.
It can be estimated as $L_{EC}=\frac{4}{3} \sigma_T c U'_{ext} n_e \langle\gamma^2\rangle V \delta^4$, where $U'_{ext}$ is the energy density of the external radiation field measured in the frame of reference of the blob, $n_e$ is the number density of the electrons, $V=4/3 \uppi R^3$ is the volume of the emission region measured in its own frame of reference and $\delta$ is the Doppler factor of the blob. 
The average squared Lorentz factor of the electrons $\langle\gamma^2\rangle$ can be approximated as $\gamma_b$ if the distribution starts from $\gamma_{min}\approx 1$ and follows an index of $2$ up to the break of $\gamma_b$.
Assuming $\Gamma\approx \delta$ the total kinetic power of a jet composed of cold protons with number density of $n_p$ is $P_{\rm jet}= \uppi R^2 n_p m_p c^3 \Gamma^2$.
Combining the two formulas, for a jet with $n_e \approx n_p$ we obtain 
$P_{\rm jet}=1.3 L_{EC} (U'_{ext}/\mathrm{0.1erg\,cm^{-3}})^{-1} (\gamma_b/10^3)^{-1} (R/10^{17}\mathrm{cm})^{-1} (\Gamma/10)^{-2}$. 
The observed luminosity of $\sim 4\times10^{47}\mathrm{erg\,s^{-1}}$ of the EC peak requires the jet Lorentz factor to be at least 10 even for the case of $P_{\rm jet}= 10 L_{\rm disk}$.
For the modeling we apply the same values as used in \cite{al14} for the jet bulk Lorentz factor $\Gamma=20$ and $\delta=25$. 
For such a bulk Lorentz factor the assumed jet semi-aperture angle of $0.047$\,rad is broader than $0.01(\Gamma/20)^{-1}$ suggested by radio observations (see e.g. \citealp{jo05}).
The radio observations are sensitive to the jet opening angle at a distance equal to or greater than the location of the radio core, i.e. a few pc from the base of the jet.
On the other hand, the emission region assumed in the modelling is located closer to the base of the jet, where the jet opening angle can be larger, as observed e.g. in the case of M87 radio galaxy \cite{an12}.

\begin{table*}
\centering
\begin{tabular}{cccccccccc}
\hline
\hline
 & $\gamma _{\rm min}$ & $\gamma _{\rm c}$ & $\gamma _{\rm b}$& $\gamma _{\rm max}$& $n_0$& $n_1$&$n_2$ &$B$ &$K$\\
& [1] & [2]  & [3] & [4] & [5] & [6] & [7] & [8] & [9] \\
\hline
Period A    &$1$ & $150$ & $800$     & $4\times 10^{4}$ & 1 & $2$ & $3.7 $      & $0.23 $&$ 3.0\times 10^{4}$\\      
Period B    & $1$      & $150$ &$500$& $3\times 10^{4}$ & 1 & $2$       & $3.7$ & $0.34$ &$ 2.6\times 10^{4}$\\      
\hline
\hline
\end{tabular}
\vskip 0.4 true cm
\caption{
Input model parameters for the models of epochs A (MJD 57160-57161) and B (57164-57166) of \src\ in Fig. \ref{fig:sed}. 
[1], [2], [3]  and [4]: minimum, cooling break, acceleration break and maximum electron Lorentz factor respectively.  
[5], [6]  and [7]: slope of the electron energy distribution below $\gamma_c$, between $\gamma_c$ and $\gamma_b$, and above $\gamma_b$. 
[8]: magnetic field [G]. 
[9]: normalization of the electron distribution in units of cm$^{-3}$.}
\label{tab:param}
\end{table*}

Having fixed these values, the free parameters of the model are only the intensity of the magnetic field $B$ and those describing the electron energy distribution. 
Hence, the observed  variability according to this scenario is caused by changes in the conditions of the plasma flowing through the shock region.
Since we assume that the emission occurs outside the intense radiation field of the BLR, the IC emission and the radiative losses of the emitting electrons are dominated by the scattering off the IR radiation field of the torus. 
As the energy density of this radiation field is relatively low, the cooling of the electrons is not very effective. 
A simple calculation shows that the Lorentz factor at which the cooling time equals the dynamical time $R/c$ is of the order of $\gamma_{\rm cool}\approx 940 R_{16}^{-1} (\Gamma/20)^{-2}$. % 600 R_{16}^{-1} (\Gamma/25)^{-2}
If we assume that the electrons are injected starting from $\gamma_{\rm inj,min}=1$, with a power law distribution with slope $n_{\rm inn}$, in equilibrium we would expect a break at $\gamma_{\rm cool}$, above which the slope of the distribution would steepen to $n_{\rm inj}+1$. However, such a break could not properly describe the SED, since the required break (estimated using the X-ray and the MAGIC slopes) is larger (even taking into account Klein-Nishina effects on the spectrum in the MAGIC energy range). 
To reproduce the SED  we therefore assume a scenario in which the electron energy distribution is a double broken power law; with a cooling break at $\gamma_c$, and a second break connected e.g. with the acceleration process at $\gamma_b$. 
The particles are injected into the emission region with a broken power-law energy distribution with slope $n'_1=1$ and $n'_2=2.7$ and break Lorentz factor $\gamma_{\rm b}$.
In equilibrium conditions, the electron energy distribution displays three power-laws with slopes $n_0=n'_1$ from $\gamma_{\rm min}$ to the cooling electron Lorentz factor $\gamma_{c}$, $n_1=n'_1+1=2$ from $\gamma_{c}$ to $\gamma_{\rm b}$ and $n_2=n'_2+1=3.7$ above $\gamma_{\rm b}$. (see e.g. \citealp{gh02}). 
Note that similar hard spectra has been also postulated in modelling of \emph{blue} blazars \citep{gh12}.

The values of these parameters required to reproduce the SEDs are reported in Table \ref{tab:param}. 
In fact, the difference in the emission between Period A and B can be explained with a relatively small change in the fit parameters, mainly a slightly stronger magnetic field and lower maximal and break energies of the electrons

The model discussed here has some caveats.
As typically happens in blazar modelling, the radio points overshoot the model line, which has a strong low energy cutoff due to synchrotron self-absorption. 
This emission is normally explained to occur from much larger regions farther along the jet. 
Moreover the variations of the optical emission occur on longer time scales than the flares observed in GeV. 
Hence, additional optical emission might be produced by the high energy electrons swept with the flow farther along the jet (up to a few pc from the base of the jet).
In those regions the external radiation field density would be too low to efficiently produce high energy photons via inverse Compton process, turning synchrotron emission into the dominant radiation process. 
In spite of adiabatic energy losses that electrons can suffer, the total observed synchrotron radiation might still slowly increase due to aggregation of electrons from multiple individual flares.
In fact, if the electrons reach the radio core (and beyond) located at the distance of $d_{\rm core}$ they could be responsible for the emission of a new radio knot. 
The separation time of such a knot from the core could be estimated to occur 
$(1+z) d_{\rm core}/ (c \delta \Gamma) = 21 (d_{\rm core}/6.5\mathrm{pc}) (\delta/25)^{-1} (\Gamma/20)^{-1}$\,days after the gamma-ray flare.
The extent of the core might shorten the time delay before the knot-core interaction starts.
We encourage further trials of modelling of the observed high state with scenarios employing emission from larger length of the jet than a single active region.

%______________________________________________________________
\section{Discussion and Conclusions}\label{sec:conc}
Using the MAGIC telescopes data we have detected enhanced VHE gamma-ray emission from the direction of \src\ during the high optical and GeV state of the source in May 2015.
It is the first time that VHE gamma-ray variability was detected for this source. 
The spectral shape is, however, consistent within the statistical uncertainties with the previous measurements of the source. 
During May 2015 the IR through UV data showed a gradual increase of the flux, while the flux in the X-ray range was slowly decreasing. 

The May 2015 data revealed, similarly to the 2012 data, that the enhanced VHE gamma-ray emission occurred during the rotation of the optical polarization angle. 
Also, similarly to other gamma-ray flares, an ejection of a new radio component was observed, however, with an unusual position angle. 
This suggests that the association of VHE gamma-ray emission with rotation of EVPA and ejection of a new radio component might be a common feature of \src .

The source was modelled with the external Compton scenario. 
The evolution of the state of the source from the VHE gamma-ray flare to a lower emission (at the level of 2012 high state reported in \citealp{al14}) can be explained by relatively small changes in the conditions of the plasma flowing through the emission region. 
The presented scenario is, however, only one possible solution. 
As discussed in \citealp{al14}, if we assume that the VHE flaring is indeed connected to the ejection of the new component (in this case $K15$) from the VLBA core and the rotation of the optical polarization angle, it would be natural to assume a single emission region located far outside the dusty torus. 
In this case the seed photons could be provided by the sheath of the jet and this scenario has been shown to provide feasible description of the previous flaring epochs of \src\ \citep{al14,md15}. 
The VHE gamma-ray variability with time scale $\tau$ seen during the 2015 outburst puts constraints on the size, and therefore also on the location of the emission region. 
Assuming that the spine of the jet fills a significant fraction of the jet (as in \citealp{al14}), the location of the emission region cannot be farther than $d=\tau \delta c / \left( (1+z)\theta_{\rm j}\right) = 2.7 (t/3\,\mathrm{days})(\delta/25)(\theta_{\rm j}/ 1^\circ)^{-1}$\,pc.
Therefore, for placing the emission region at the radio core a high Doppler factor and a narrow jet are needed. 
In fact such low values of the jet extension, $(0.2\pm0.2)^\circ$ \citep{jo05} and $0.9^\circ$ \citep{pu09} at the radio core are reported by the radio observations.
Alternatively, the inner spine can be much narrower than the whole jet, as suggested by \cite{md15}.
 
To further study the connection of VHE emission with events in lower frequencies, long-term monitoring data are needed and this question will be addressed in a future publication.
With the detection of this flare from \src , VHE gamma-ray variability (on times scales varying from tens of minutes to days) has been observed in all FSRQs known in VHE gamma rays. 
Fast-varying VHE gamma-ray emission is common among the brightest gamma-ray FSRQs. 
As it seems that most of the gamma-ray FSRQs can only be detected during these flares, it is indeed not surprising that only a handful have been detected in VHE gamma-rays.

\begin{acknowledgements}
%MAGIC
The MAGIC collaboration would like to thank
the Instituto de Astrof\'{\i}sica de Canarias
for the excellent working conditions
at the Observatorio del Roque de los Muchachos in La Palma.
The financial support of the German BMBF and MPG,
the Italian INFN and INAF,
the Swiss National Fund SNF,
the he ERDF under the Spanish MINECO
(FPA2015-69818-P, FPA2012-36668, FPA2015-68278-P,
FPA2015-69210-C6-2-R, FPA2015-69210-C6-4-R,
FPA2015-69210-C6-6-R, AYA2013-47447-C3-1-P,
AYA2015-71042-P, ESP2015-71662-C2-2-P, CSD2009-00064),
and the Japanese JSPS and MEXT
is gratefully acknowledged.
This work was also supported
by the Spanish Centro de Excelencia ``Severo Ochoa''
SEV-2012-0234 and SEV-2015-0548,
and Unidad de Excelencia ``Mar\'{\i}a de Maeztu'' MDM-2014-0369,
by grant 268740 of the Academy of Finland,
by the Croatian Science Foundation (HrZZ) Project 09/176
and the University of Rijeka Project 13.12.1.3.02,
by the DFG Collaborative Research Centers SFB823/C4 and SFB876/C3,
and by the Polish MNiSzW grant 745/N-HESS-MAGIC/2010/0.
%Fermi
The \textit{Fermi} LAT Collaboration acknowledges generous ongoing support
from a number of agencies and institutes that have supported both the
development and the operation of the LAT as well as scientific data analysis.
These include the National Aeronautics and Space Administration and the
Department of Energy in the United States, the Commissariat \`a l'Energie Atomique
and the Centre National de la Recherche Scientifique / Institut National de Physique
Nucl\'eaire et de Physique des Particules in France, the Agenzia Spaziale Italiana
and the Istituto Nazionale di Fisica Nucleare in Italy, the Ministry of Education,
Culture, Sports, Science and Technology (MEXT), High Energy Accelerator Research
Organization (KEK) and Japan Aerospace Exploration Agency (JAXA) in Japan, and
the K.~A.~Wallenberg Foundation, the Swedish Research Council and the
Swedish National Space Board in Sweden.
%Metsahovi
The Mets\"ahovi team acknowledges the support from the Academy of Finland to our observing projects (numbers 212656, 210338, 121148, and others).
%SMARTS
This paper has made use of up-to-date SMARTS optical/near-infrared light curves that are available at \url{www.astro.yale.edu/smarts/glast/home.php}.
%NICS
MIRO is operated by Physical Research Laboratory, Ahmedabad with support from Dept of Space, government of India.
% TCS
This article is based on observations made with the 1.5~m telescope Carlos S\'anchez operated by the Instituto de Astrofisica de Canarias in the Teide Observatory. 
% polarimetry
The data presented here were obtained in part with ALFOSC, which is provided by the Instituto de Astrofisica de Andalucia (IAA) under a joint agreement with the University of Copenhagen and NOTSA.
%RINGO3
The Liverpool Telescope is operated on the island of La Palma by Liverpool John Moores University in the Spanish Observatorio del Roque de los Muchachos of the Instituto de Astrofisica de Canarias with financial support from the UK Science and Technology Facilities Council.
%AZT & LX-200
St.Petersburg University team acknowledges support from Russian RFBR grant 15-02-00949 and St.Petersburg University research grant 6.38.335.2015.
%VLBA
The BU group acknowledges support by NASA under {\it Fermi} Guest
Investigator grants NNX11AQ03G and NNX14AQ58G.
The VLBA is an instrument of the National Radio Astronomy Observatory. The National Radio Astronomy Observatory is a facility of the National Science
Foundation operated under cooperative agreement by Associated Universities, Inc.
We would like to thank the anonymous journal referee for the comments on the manuscript.
\end{acknowledgements}

%-------------------------------------------------------------------


\begin{thebibliography}{}
\bibitem[Abdo et al.(2010)]{ab10} Abdo, A.~A., Ackermann, M., Agudo, I., et al.\ 2010, \apj, 721, 1425 % Fermi 2008-2009 flares 
\bibitem[Abramowski et al.(2013)]{ab13} H.E.S.S.~Collaboration, Abramowski, A., Acero, F., et al.\ 2013, \aap, 554, A107 % HESS detection
\bibitem[Ackermann et al.(2011)]{ac11} Ackermann, M., Ajello, M., Allafort, A., et al.\ 2011, \apj, 743, 171 
\bibitem[Ahnen et al.(2015)]{ah15} Ahnen, M.~L., Ansoldi, S., Antonelli, L.~A., et al.\ 2015, \apjl, 815, L23 % PKS1441
\bibitem[Acero et al.(2015)]{Acero15} Acero, F., Ackermann, M., Ajello, M. et al. 2015, ApJS, 218, 23 % 3FGL
\bibitem[Acero et al.(2016)]{ac16} Acero, F., Ackermann, M., Ajello, M., et al.\ 2016, \apjs, 223, 26 % Fermi diffuse model
\bibitem[Albert et al.(2007)]{al07} Albert, J., Aliu, E., Anderhub, H., et al.\ 2007, NIM A, 583, 494 % Nuclear Instruments and Methods in Physics Research A, unfolding
\bibitem[Albert et al.(2008)]{al08} MAGIC Collaboration, Albert, J., Aliu, E., et al.\ 2008, Science, 320, 1752 % 3c279
\bibitem[Aleksi{\'c} et al.(2011)]{al11} Aleksi{\'c}, J., Antonelli, L.~A., Antoranz, P., et al.\ 2011, \apjl, 730, L8 % PKS1222
\bibitem[Aleksi{\'c} et al.(2014)]{al14} Aleksi{\'c}, J., Ansoldi, S., Antonelli, L.~A., et al.\ 2014, \aap, 569, A46 %  PKS1510, MAGIC
\bibitem[Aleksi{\'c} et al.(2016a)]{al16a} Aleksi{\'c}, J., Ansoldi, S., Antonelli, L.~A., et al.\ 2016, Astroparticle Physics, 72, 61 
\bibitem[Aleksi{\'c} et al.(2016b)]{al16b} Aleksi{\'c}, J., Ansoldi, S., Antonelli, L.~A., et al.\ 2016, Astroparticle Physics, 72, 76 % upgrade, part 2
\bibitem[Arnold et al.(2012)]{ar12} Arnold, D.~M., Steele, I.~A., Bates, S.~D., Mottram, C.~J., \& Smith, R.~J.\ 2012, \procspie, 8446, 84462J
\bibitem[Asada \& Nakamura(2012)]{an12} Asada, K., \& Nakamura, M.\ 2012, \apjl, 745, L28
\bibitem[Atwood et al.(2009)]{Atwood09} Atwood, W. B., Abdo, A. A., Ackermann, M., et al. 2009, ApJ, 697, 1071
\bibitem[Banerjee \& Ashok(2012)]{ba12} Banerjee, D.~P.~K., \& Ashok, N.~M.\ 2012, Bulletin of the Astronomical Society of India, 40, 243 
\bibitem[Bentz et al.(2009)]{be09} Bentz, M.~C., Peterson, B.~M., Netzer, H., Pogge, R.~W., \& Vestergaard, M.\ 2009, \apj, 697, 160
\bibitem[Bessell et al.(1998)]{be98} Bessell, M.~S., Castelli, F., \& Plez, B.\ 1998, \aap, 333, 231 % bands to flux conversion
\bibitem[Bonning et al.(2012)]{bo12} Bonning, E., Urry, C.~M., Bailyn, C., et al.\ 2012, \apj, 756, 13
\bibitem[Burrows et al.(2004)]{2004SPIE.5165..201B} Burrows, D.~N., Hill, J.~E., Nousek, J.~A., et al.\ 2004, \procspie, 5165, 201 
\bibitem[Carrasco et al.(2015)]{atel7804} Carrasco, L., Porras, A., Recillas, E., et al.\ 2015, The Astronomer's Telegram, \#7804 
\bibitem[Cawthorne et al.(2013)]{ca13} Cawthorne, T.~V., Jorstad, S.~G., \& Marscher, A.~P.\ 2013, \apj, 772, 14
\bibitem[Clarke \& Neumayer(2002)]{cn02} Clarke, D., \& Neumayer, D.\ 2002, \aap, 383, 360
\bibitem[Cutri et al.(2003)]{cu03} Cutri, R.~M., Skrutskie, M.~F., van Dyk, S., et al.\ 2003, The IRSA 2MASS All-Sky Point Source Catalog, NASA/IPAC Infrared Science Archive. \burl{http://irsa.ipac.caltech.edu/applications/Gator/}  
\bibitem[Chatterjee et al.(2008)]{chatterjee08} Chatterjee, R., Jorstad, S. G., Marscher, A. P. et al. 2008, ApJ, 689, 79
\bibitem[D'Arcangelo et al.(2007)]{da07} D'Arcangelo, F.~D., Marscher, A.~P., Jorstad, S.~G., et al.\ 2007, \apjl, 659, L107
\bibitem[D'arcangelo et al.(2009)]{da09} D'arcangelo, F.~D., Marscher, A.~P., Jorstad, S.~G., et al.\ 2009, \apj, 697, 985
\bibitem[Dom{\'{\i}}nguez et al.(2011)]{do11} Dom{\'{\i}}nguez, A., Primack, J.~R., Rosario, D.~J., et al.\ 2011, \mnras, 410, 2556 % EBL model
\bibitem[Dotson et al.(2015)]{do15} Dotson, A., Georganopoulos, M., Meyer, E.~T., \& McCann, K.\ 2015, \apj, 809, 164 % location of 2009 flares
\bibitem[Fruck \& Gaug(2015)]{fg15} Fruck, C., \& Gaug, M.\ 2015, European Physical Journal Web of Conferences, 89, 02003 % LIDAR
\bibitem[Ghisellini \& Tavecchio(2009)]{gt09} Ghisellini, G., \& Tavecchio, F.\ 2009, \mnras, 397, 985
\bibitem[Ghisellini et al.(2002)]{gh02} Ghisellini, G., Celotti, A., \& Costamante, L.\ 2002, \aap, 386, 833
\bibitem[Ghisellini et al.(2010)]{gh10} Ghisellini, G., Tavecchio, F., Foschini, L., et al.\ 2010, \mnras, 402, 497 % properties of bright Fermi blazars
\bibitem[Ghisellini et al.(2012)]{gh12} Ghisellini, G., Tavecchio, F., Foschini, L., et al.\ 2012, \mnras, 425, 1371 % hard spectra
\bibitem[Hovatta et al.(2016)]{hovatta16} Hovatta T., Lindfors, E., Pavlidou, V. et al. 2016, A\&A, submitted
\bibitem[Jankowsky et al.(2015)]{atel7799} Jankowsky, F., Zacharias, M., Wierzcholska, A., et al.\ 2015, The Astronomer's Telegram, \# 7799 % optical ATOM
\bibitem[Jermak et al.(2016)]{je16} Jermak, H., Steele, I.~A., Lindfors, E., et al.\ 2016, \mnras, 462, 4267
\bibitem[Jorstad et al.(2005)]{jo05} Jorstad, S.~G., Marscher, A.~P., Lister, M.~L., et al.\ 2005, \aj, 130, 1418 
\bibitem[Kalberla et al.(2005)]{2005A&A...440..775K} Kalberla, P.~M.~W., Burton, W.~B., Hartmann, D., et al.\ 2005, \aap, 440, 775 
\bibitem[Larionov et al.(2008)]{la08} Larionov, V.~M., Jorstad, S.~G., Marscher, A.~P., et al.\ 2008, \aap, 492, 389
\bibitem[MacDonald et al.(2015)]{md15} MacDonald, N.~R., Marscher, A.~P., Jorstad, S.~G., \& Joshi, M.\ 2015, \apj, 804, 111 
\bibitem[Marscher et al.(2010)]{ma10} Marscher, A.~P., Jorstad, S.~G., Larionov, V.~M., et al.\ 2010, \apjl, 710, L126 
\bibitem[Mattox et al.(1996)]{ma96} Mattox J. R. et al., 1996, ApJ, 461, 396 % TS in Fermi
\bibitem[Mirzoyan(2015)]{atel7542} Mirzoyan, R. for the  MAGIC Collaboration 2015, The Astronomer's Telegram, \#7542
\bibitem[Mirzoyan(2016)]{atel9105} Mirzoyan, R.\ 2016, The Astronomer's Telegram, \#9105,  
\bibitem[de Naurois(2016)]{atel9102} de Naurois, M.\ 2016, The Astronomer's Telegram, \#9102,  
\bibitem[Poole et al.(2008)]{po08} Poole, T.~S., Breeveld, A.~A., Page, M.~J., et al.\ 2008, \mnras, 383, 627 
\bibitem[Pushkarev et al.(2009)]{pu09} Pushkarev, A.~B., Kovalev, Y.~Y., Lister, M.~L., \& Savolainen, T.\ 2009, \aap, 507, L33 % jet opening angle 
\bibitem[Pushkarev et al.(2012)]{pu12} Pushkarev, A.~B., Hovatta, T., Kovalev, Y.~Y., et al.\ 2012, \aap, 545, A113
\bibitem[Raiteri et al.(2010)]{ra10} Raiteri, C.~M., Villata, M., Bruschini, L., et al.\ 2010, \aap, 524, A43
\bibitem[Rolke et al.(2005)]{ro05} Rolke, W.~A., L{\'o}pez, A.~M., \& Conrad, J.\ 2005, NIM A, 551, 493 % ULs
\bibitem[Saito et al.(2013)]{sa13} Saito, S., Stawarz, {\L}., Tanaka, Y.~T., et al.\ 2013, \apjl, 766, L11 % PKS1510 in Fermi, flares
\bibitem[Saito et al.(2015)]{sa15} Saito, S., Stawarz, {\L}., Tanaka, Y.~T., et al.\ 2015, \apj, 809, 171 
\bibitem[Sameer et al.(2015)]{atel7495} Sameer, Kaur, N., Ganesh, S., Kumar, V., \& Baliyan, K.~S.\ 2015, The Astronomer's Telegram, \#7495 % MIRO IR ATel
\bibitem[Schlafy \& Finkbeiner (2011)]{schlafy11} {Schlafly}, E.~F. and {Finkbeiner}, D.~P., 2011, ApJ, 737, 103S
\bibitem[Sikora et al.(1994)]{sbr94} Sikora, M., Begelman, M.~C., \& Rees, M.~J.\ 1994, \apj, 421, 153 
\bibitem[Smith et al.(2009)]{smith} Smith, P.S., Montiel, E., Rightley, S., Turner, J., Schmidt, G.D., \& Jannuzi, B.T. 2009, arXiv:0912.3621, 2009 Fermi Symposium, eConf Proceedings C091122
\bibitem[Steele et al.(2004)]{st04} Steele, I.~A., Smith, R.~J., Rees, P.~C., et al.\ 2004, \procspie, 5489, 679 
\bibitem[Tanner et al.(1996)]{ta96} Tanner, A.~M., Bechtold, J., Walker, C.~E., Black, J.~H., \& Cutri, R.~M.\ 1996, \aj, 112, 62 % redshift
\bibitem[Ter\"aesranta et al.(1998)]{te98} Tera\"esranta, H., Tornikoski, M., Mujunen, A., et al.\ 1998, \aaps, 132, 305 %Metsahovi
\bibitem[Villata et al.(1997)]{villata97} Villata M., Raiteri, C. M., Ghisellini, G. et al. 1997 A\&AS 121, 119
\bibitem[Zacharias et al.(2016)]{za16} Zacharias, M., B{\"o}ttcher, M., Chakraborty, N., et al.\ 2016, arXiv:1611.02098
\bibitem[Zanin et al.(2013)]{za13} Zanin, R., Carmona, E., Sitarek, J., et al., 2013, Proc of 33rd ICRC, Rio de Janeiro, Brazil, Id. 773 % MARS
\bibitem[Zhang et al.(2015)]{zh15} Zhang, H., Chen, X., B{\"o}ttcher, M., Guo, F., \& Li, H.\ 2015, \apj, 804, 58
\end{thebibliography}
\end{document}